\documentclass[
    10pt,
    aps,
    prb,
    twocolumn,
    superscriptaddress
]{revtex4}

\usepackage{graphicx} 
\usepackage{epsfig}
\usepackage{dcolumn}
\usepackage{bm}
\usepackage{amsmath}
\usepackage{amsfonts}
\usepackage{latexsym}
\usepackage{amssymb}
\usepackage{color}
\usepackage{hyperref}
\usepackage[normalem]{ulem}
\usepackage{subfig}
\usepackage{tasks}
\usepackage{float}

\newcommand\upermat[2]{%
  \makebox[0pt][l]{$\smash{\overbrace{\phantom{%
    \begin{array}{cccc}#2 & \hphantom{a}\end{array}}}^{\text{$#1$}}}$}#2}

\newcommand\undermat[2]{%
  \makebox[0pt][l]{$\smash{\underbrace{\phantom{%
    \begin{array}{cccc}#2 & \hphantom{a}\end{array}}}_{\text{$#1$}}}$}#2}

\usepackage{tikz,xcolor,hyperref}
\definecolor{lime}{HTML}{A6CE39}
\DeclareRobustCommand{\orcidicon}{%
	\begin{tikzpicture}
	\draw[lime, fill=lime] (0,0)
	circle [radius=0.16]
	node[white] {{\fontfamily{qag}\selectfont \tiny ID}};
	\draw[white, fill=white] (-0.0625,0.095)
	circle [radius=0.007];
	\end{tikzpicture}
	\hspace{-2mm}
}
\foreach \x in {A, ..., Z}{%
	\expandafter\xdef\csname orcid\x\endcsname{\noexpand\href{https://orcid.org/\csname orcidauthor\x\endcsname}{\noexpand\orcidicon}}
}

\begin{document}
\title{Fate of gapless edge states in two-dimensional topological insulators with Hatsugai-Kohmoto interaction}
\author{Jan Skolimowski\orcidA}
\email{jskolimowski@magtop.ifpan.edu.pl}
\affiliation{International Research Centre MagTop, Institute of
   Physics, Polish Academy of Sciences,\\ Aleja Lotnik\'ow 32/46,
   PL-02668 Warsaw, Poland}
\author{Wojciech Brzezicki\orcidB}
\email{brzezicki@magtop.ifpan.edu.pl}
\affiliation{Institute of Theoretical Physics, Jagiellonian University, Prof. S. \L{}ojasiewicza 11, PL-30348 Krak\'ow, Poland}
\affiliation{International Research Centre MagTop, Institute of
   Physics, Polish Academy of Sciences,\\ Aleja Lotnik\'ow 32/46,
   PL-02668 Warsaw, Poland}

\date{\today}

\begin{abstract}
   Topologically protected edge states are the highlight feature of an interface between non-equivalent insulators. The robustness/sensitivity of these states to local single-particle perturbations is well understood, while their stability in the presence of various types of two-particle interactions remains unclear. To add to previous discussions of the Hubbard and unscreened Coulomb interactions, we address this problem from the point of view of infinite-range Hatsugai-Kohmoto interaction. Based on our numerical results for two models of Chern insulators, the Kane-Mele and spinful Haldane model, on a ribbon geometry with zig-zag edges, we argue that any finite interaction strength $U$ is sufficient to open a charge gap in the spectrum of either Chern insulator. We explain the differences between the two cases and present how their edge states phase out as the system enters the strongly correlated phase. We show that the closing of the many-body gap in periodic variants of these models can be connected to the onset of hybridization between the edge and bulk modes in finite geometries. Providing an example of the {\it bulk-boundary correspondence} in systems where there is a topological phase transition without closing of the spectral gap.
\end{abstract}

\maketitle
\section{Introduction}

The notion of special states at the interface of two topologically non-equivalent insulators has been well established, starting from the early works by Halperin on the integer quantum Hall state\cite{PhysRevB.25.2185} to later works by Haldane\cite{PhysRevLett.61.2015}, Kane and Mele\cite{PhysRevLett.95.146802} and many others. These states called the edge states (ES), cross the bulk band gap and are the smoking gun for the existence of insulators topologically different than the vacuum state. This is known as the {\it bulk-boundary correspondence}. The ES have a topological protection stemming from the symmetry of the bulk crystal, which provides them with properties impossible to realize by trivial surface physics. Because they are a topological feature they are insensitive to small perturbations that do not break the symmetry of the bulk. In the case of quantum spin Hall (QSH) state this symmetry is the time-reversal symmetry that protects the helical ES from back-scattering, opening doors for dissipationless spin-current. 
Thus these states are not only important from the point of view of fundamental studies, as they provide a clear indication of non-trivial topology, but due to their unusual properties, guaranteed by the topological protection, they are a promising platform for application in i.e. spintronics\cite{Niyazov2020}.

Almost all the concepts for analyzing the topology of a system, developed using the single-particle picture, start to break down once the correlations are taken into account. Not only the extension of the ten-fold way\cite{Zirnbauer1, Zirnbauer2} is missing but there is even no consensus on a clear indicator of a topological phase transition (TPT) in many-body systems\cite{PhysRevX.2.031008,PhysRevX.4.011006, PhysRevLett.105.256803, PhysRevB.108.125115,Wagner2023}. The recent reports show that even the celebrated gap closure across the TPT does not have to take place\cite{PhysRevB.108.035121,PhysRevB.110.075105}. The long-range effects caused by interactions make one also question the applicability of the {\it bulk-boundary correpsondece} to many-body systems. For the QSH state and local interactions, it was initially shown that the ES are robust against weak local interactions \cite{PhysRevB.73.045322,PhysRevB.79.235321} but later this claim was questioned\cite{PhysRevB.90.035116}. Nonetheless, both reports agree that in certain cases interaction will change the charge-conducting edge channel into a charge-gaped but spin-conducting state. A similar state was reported for unscreened Coulomb interactions\cite{PhysRevLett.101.196804}.

The issue with studying the topology of many-body systems originates from a lack of general solutions of the established models of correlations, i.e. the Hubbard model, and scarcity of solvable toy models that can grasp the crucial behaviors of it. Recently, the model proposed by Hatsugai-Kohmoto has been put forward as a proxy for the rise of Mott phase\cite{PhillipsFixedPoint}, one of the key features of the Hubbard model. It is an exactly solvable model, in the thermodynamic limit, which immediately drew a lot of attention from the solid-state community\cite{PhysRevB.106.155119, PhysRevB.108.085106, Zhao_2023, PhillipsSC,guerci2024, zhao2024topicreviewhatsugaikohmotomodels}. Multiple studies analyzed some version of this interaction model in the context of the interplay between correlations and topology\cite{Mai2023,PhysRevB.108.195145,PhysRevB.110.075105}. All of them did so in the canonical for this model momentum space basis, which demands the translation invariance of the model. Hence the fate of the {\it bulk-boundary correspondence} in systems with H-K interaction was not explored. The usual reasoning that if the bulk model has a non-zero invariant then its interface with trivial insulator has to host gapless boundary modes is especially challenged in the presence of the H-K interaction, as its form was shown to be altered upon changing the boundary conditions \cite{PhysRevB.109.165129}. Here we answer the looming question about the existence and stability of gapless edge modes in two-dimensional topological insulators (2DTI) with H-K interaction. For that purpose we focus on the two prototypical models of 2DTI characterized by a different topological invariant: (i) the spinful Haldane model (SHM)\cite{PhysRevB.83.205116,PhysRevB.88.241101} and (ii) the Kane-Mele model (KMM)\cite{PhysRevLett.95.146802,PhysRevLett.95.226801} and consider them on a zigzag graphene nanoribbon (ZGNR)\cite{Son2006,PhysRevLett.95.226801,PhysRevLett.95.146802} geometry, that is known to induce boundary modes. To discuss the {\it bulk-boundary correspondence} we will also present the results for the corresponding bulk variants of these two models. The previous studies on KMM with H-K interactions focused mainly on the H-K interaction in the orbital basis, thus it is necessary to quantify the role played by the effective inter-sublattice couplings, that H-K interaction can introduce

Our main finding is, that the long-range nature of the Hatsugai-Kohmoto interaction is detrimental to the existence of gapless edge modes in either model of a topological insulator. The numerical results hint that any finite interaction strength leads to the opening of a charge gap at the Fermi level, that only increases with $U$. Yet, the localized nature of the edge states survives until a much larger interaction strength. With the help of effective models, we explain the differences observed between the two models and connect them to the dominant inter-edge scattering processes allowed by the undelaying lattice symmetries. The accuracy of the effective description provides more arguments in favor of the formation of the charge gap in the edge modes at any finite $U$. Following similar studies of different interaction types, we show that gapless spin edge modes can survive the onset of H-K interaction, with their robustness depending on the symmetry of the 2DTI. Analyzing the periodic variants of the two models in the thermodynamic limit we show different behaviors of their topological invariant. Most noticeably the inter-sublattice interaction terms secure a non-degenerate ground-state, adiabatically connected to the $U=0 $ QSH state, around the $K/K^\prime$ points 
for any $U$ in the KMM, while the Chern insulating state of the SHM ceases to exist above a certain critical $U_c$. 
Lastly, comparing the results for a ZGNR and a full translation invariant lattice we link the interaction strength $U$ at which the latter starts to have points in the Brillouin zone with a degenerate ground state, the breakdown of the integer topological invariant, with the crossing of the edge and bulk modes at the $k_x=\pi$ in the former, mixed bulk-edge nature of the low-energy modes. This gives a new perspective on the {\it bulk-boundary correspondence} in systems where the TPT is not accompanied by the closing of the spectral gap.

\section{Models}
\begin{figure}
    \centering
    \includegraphics[width=\linewidth]{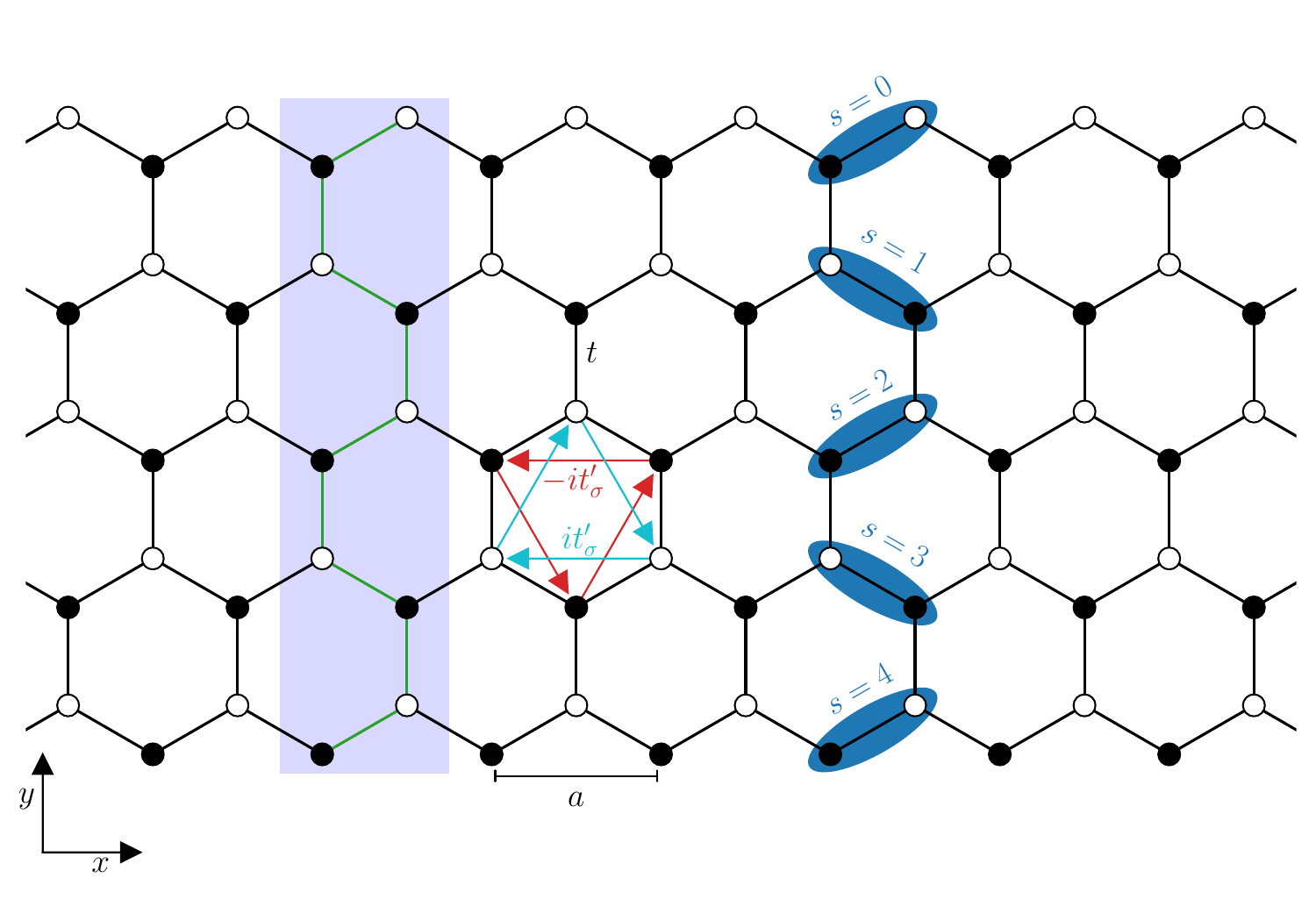}
    \caption{Schematic depiction of the graphene zig-zag nanoribbon with 10 sites across, shared by the KMM and the SHM. Black lines are the inter-sublattice couplings $t$, colored arrows are the complex next-nearest neighbor couplings within a single hexagon. The blue-shaded region is the supercell of the ZGNR, and the green line highlights the armchair-shaped path across the ribbon. The relative shift along $x$ between the sites closer to the right and to the left edge of the shaded area is the reason why an additional gauge transformation $f(y_1,y_2,y_3,y_4)$ is introduced in Eq.\ref{HK_mixed}. The blue ellipses depict the pairs of sites from each sublattice which are enumerated by the $s$-index. 
    }
    \label{lattice_pic}
\end{figure}

Models studied in this work are geometrically similar since they share a common ancestor - the spinless Haldane model\cite{PhysRevLett.61.2015}. Both describe electrons on a honeycomb lattice with a complex next-nearest neighbor whose sign depends on the sublattice, depicted in Fig. \ref{lattice_pic}. Despite that, they are characterized by a different topological invariant in the non-interacting limit. The SHM, being a direct sum of the Haldane models for each spin channel, inherits the broken time-reversal symmetry and is described by a Chern number that can go up to the number of spin channels. In the KMM the time-reversal symmetry is restored by the addition of spin-dependent sign of the complex next-nearest neighbor hopping. In that case, the Chern number from both spin channels cancels out. The proper topological invariant to characterize the KMM is the spin-Chern number. The kinetic part of the Hamiltonian describing KMM is given by
\begin{multline}
    \mathcal{H}_{KMM}=-t\sum_{\langle \gamma,\gamma^\prime\rangle, \sigma}c^{\dagger}_{\gamma,\sigma}c_{\gamma^\prime,\sigma}\\+it^\prime\sum_{\langle \langle \gamma,\gamma^\prime\rangle \rangle, \sigma, \sigma^\prime}\nu_{\gamma,\gamma^\prime}c^\dagger_{\gamma,\sigma}s^z_{\sigma,\sigma^\prime}c_{\gamma^\prime,\sigma^\prime},
\end{multline}
where $c_{\gamma,\sigma}$ is the fermionic annihilation operator at site $\gamma$ and with spin $\sigma$. The nearest and next-nearest neighbor hopping strengths are controlled by $t$ and $t^\prime$ respectively. The restricted summations $\langle \gamma,\gamma^\prime\rangle$ and $\langle \langle \gamma,\gamma^\prime\rangle \rangle$ reflect the range of these hoppings which are the nearest and next-nearest neighbor, respectively. The parameter $\nu_{\gamma,\gamma^\prime}=\pm 1$ distinguishes between a clockwise and anticlockwise movement of particles within either sublattice. The additional spin-dependence of this purely imaginary hopping is introduced by the Pauli z-matrix $s^z_{\sigma,\sigma^\prime}$. Hence, the corresponding Hamiltonian for the SHM lacks it and is given by  
\begin{multline}
    \mathcal{H}_{SHM}=-t\sum_{\langle \gamma,\gamma^\prime\rangle, \sigma}c^{\dagger}_{\gamma,\sigma}c_{\gamma^\prime,\sigma}\\+it^\prime\sum_{\langle \langle \gamma,\gamma^\prime\rangle \rangle, \sigma}\nu_{\gamma,\gamma^\prime}c^\dagger_{\gamma,\sigma}c_{\gamma^\prime,\sigma}.  
\end{multline}
Because the graphene lattice has a two-site unit cell, in the text below we will differentiate them by substituting the generic annihilation operator $c_{\gamma,\sigma}$ with a pair of annihilation operators $a_{\gamma,\sigma},\: b_{\gamma,\sigma}$. In this way, index $\gamma$ enumerates unit cells and the two operators span the Hilbert space of each sublattice. 

The Hatsugai-Kohmoto interaction is defined as an infinite-range two-particle interaction, which conserves the center-of-mass (COM)\cite{HK}. In its original real-space form it is given by 
\begin{equation}
    \mathcal{H}_{H-K}=\frac{U}{L^d}\sum_{j_1,j_2,j_3,j_4}\delta_{j_1+j_3,j_2+j_4} c^\dagger_{j_1,\uparrow}c_{j_2,\uparrow}c^\dagger_{j_3,\downarrow}c_{j_4,\downarrow},
\end{equation}
 where $L^d$ is the number of sites within the lattice ($d$ is the dimension) and $j_i$ are multi-indices that specify the location of a certain site. In the case of the two sites per unit cell and using the two fermionic operator basis, for each sublattice separately, it can be rewritten as 
 \begin{multline}
    \mathcal{H}_{H-K}=\frac{U}{(2 N)^d}\sum_{j_1,j_2,j_3,j_4}\delta_{j_1+j_3,j_2+j_4}\times\\
    \left(a^\dagger_{j_1,\uparrow}a_{j_2,\uparrow}a^\dagger_{j_3,\downarrow}a_{j_4,\downarrow}+b^\dagger_{j_1,\uparrow}b_{j_2,\uparrow}b^\dagger_{j_3,\downarrow}b_{j_4,\downarrow}+\right.\\
    \left(a^\dagger_{j_1,\uparrow}a_{j_2,\uparrow}b^\dagger_{j_3,\downarrow}b_{j_4,\downarrow}+b^\dagger_{j_1,\uparrow}b_{j_2,\uparrow}a^\dagger_{j_3,\downarrow}a_{j_4,\downarrow}+\right.\\
    \left.+a^\dagger_{j_1,\uparrow}b_{j_2,\uparrow}b^\dagger_{j_3,\downarrow}a_{j_4,\downarrow}+b^\dagger_{j_1,\uparrow}a_{j_2,\uparrow}a^\dagger_{j_3,\downarrow}b_{j_4,\downarrow}\right)
 \end{multline}
The indices $j_i$, now, enumerate the unit cells, thus $U$ is divided by $(2 N)^d$, which is the number of sites for an $N$ unit cell system. The conservation of COM for the inter-unit cell couplings is secured by the Kronecker-$\delta$ while for the inter-sublattice couplings, the same is obtained by explicitly writing the allowed terms. 
The interaction terms are separated into two groups: processes conserving the total spin in each sublattice (first and second row) and the spin exchange between sublattices (last row). 
 
Most studies dedicated to this model, in the translational invariant case, have not explicitly considered the latter and focused mainly on the intra-sublattice terms. There is no physical reason to neglect the inter-sublattice terms, as by construction SHM and KMM both have a single Wannier orbital but a two-sublattice structure. Thus, to establish reference points for the analysis of bulk-boundary correspondence, we will analyze the topological phase transitions that take place in the periodic versions of both models.

For the ZGNR geometry, without loss of generality, we consider edges to be along the Cartesian $x$ direction, cf. Fig. \ref{lattice_pic}. The translation invariance along the edges allows for the introduction of a Bloch plane-wave basis along $x$ characterized by a crystal momentum $k_x$. For the other variable, describing the real-space location perpendicular to the edge, we use the Cartesian $y$ coordinate. In these coordinates the kinetic terms of the Hamiltonian can be made real by a special gauge transformation\cite{PhysRevB.107.205408}, which is the consequence of the armchair shape of the cross-section of the ribbon, shown as a green line in Fig.\ref{lattice_pic}. The result of this gauge transformation is that some of the interaction terms become complex and hence the overall Hamiltonian is also complex, as we will discuss later.

In the mixed-basis representation, the single particle Hamiltonian of either SHM or KMM becomes block-diagonal in $k_x$ and spin projection $\sigma$ basis. In addition, each $k_x,\sigma$ block has a block-tridiagonal form\cite{PhysRevB.106.054209,PhysRevB.107.205408} build out of $2\times 2$ matrices. The diagonal blocks are given by:
\begin{equation}
[\mathbf{H}(k_x,\sigma)]_{s,s}=\left[
\begin{array}{cc}
-2t_\sigma ^\prime\sin(k_x)  & 2t\cos(\frac{k_x}{2})\\
2t\cos(\frac{k_x}{2}) & 2t_\sigma ^\prime\sin(k_x)
\end{array}
\right],
\end{equation}
and the off-diagonal ones are:
\begin{equation}
[\mathbf{H}(k_x,\sigma)]_{s,s+1}=\left[
\begin{array}{cc}
2 t_\sigma ^\prime \sin(\frac{k_x}{2}) & 0 \\
t & -2 t_\sigma ^\prime\sin(\frac{k_x}{2})
\end{array}
\right].
\end{equation}
and their Hermitian conjugates. The index $s\in\{0,N-1\}$ enumerates the unit cells across the ribbon, cf. Fig. \ref{lattice_pic}. The next-nearest-neighbor hopping amplitude is given by $t^\prime_\sigma$. For the time-reversal Kane-Mele model $t_\sigma ^\prime={\rm sgn}(\sigma)t^\prime$ and for the spinful Haldane model $t_\sigma^\prime=t^\prime$ and does not depend on spin orientation.

In the same mixed basis of crystal momentum $k_x$ and site index along $y$-direction the H-K interaction is given by 
\begin{multline}\label{HK_mixed}
\mathcal{H}_{H-K}= \\ \frac{U}{4N}\sum_{y_1,\ldots,y_4}\delta_{y_1+y_3,y_2+y_4}\sum_k e^{-\frac{i}{2}(y_1+y_3-y_2-y_4)k_x} \\ 
f(y_1,y_2,y_3,y_4)\sum_{\alpha=a,b}\\\left(\alpha^\dagger_{\{k_x,y_1\},\uparrow}\alpha_{\{k_x,y_2\},\uparrow}\alpha^\dagger_{\{k_x,y_3\},\downarrow}\alpha_{\{k_x,y_4\},\downarrow}  \right.\\
\left.+\alpha^\dagger_{\{k_x,y_1\},\uparrow}\alpha_{\{k_x,y_2\},\uparrow}\bar{\alpha}^\dagger_{\{k_x,y_3\},\downarrow}\bar{\alpha}_{\{k_x,y_4\},\downarrow}\right.\\
+\left. \alpha^\dagger_{\{k_x,y_1\},\uparrow}\bar{\alpha}_{\{k_x,y_2\},\uparrow}\bar{\alpha}^\dagger_{\{k_x,y_3\},\downarrow}\alpha_{\{k_x,y_4\},\downarrow} \right),
\end{multline}
where $\bar{\alpha}$ symbols the adjacent sub-lattice to $\alpha$. The function $f(y_1,y_2,y_3,y_4)$ accounts for the gauge transformation $\alpha_{\{k_x,y\}}\rightarrow e^{ik_x/2}\alpha_{\{k_x,y\}}$ on the sites which are on the right-hand side of the armchair cross-section of the ribbon, cf. green line in Fig \ref{lattice_pic}, that makes the kinetic terms real. 
The exact steps that lead to this form of interaction are given in the Appendix \ref{HK_mixed_basis_form}. 
The overall form of the interaction is the same as in the one-dimensional case, with the addition of the $k_y$ index in the fermionic operators and the phase factor introduced by the gauge transformation.

The Hamiltonian \ref{HK_mixed} does not couple sectors with different $k_x$, in contrast to standard local in-space interactions. It couples the two opposite spin channels and also introduces (correlated) hopping beyond the next-nearest neighbors, which in general can be up to the width of the ribbon. Analysis of how the latter is influencing the in-gap edge states present in the non-interacting versions of the two models is the central point of this work.

To visualize the fate of the edge states we use the local density of states $A_{\gamma,\sigma}(\omega,k_x)$ 
defined as the imaginary part of the local retarded Green's function for each of the decoupled $k_x$ block
\begin{multline}\label{Lehman}
    G^{R}_{\gamma,\sigma}(\omega,k_x)=\\
    \frac{1}{Z}\sum_{n,n^\prime}\frac{\left<n(k_x)\right|\alpha_{\gamma,\sigma} \left|n^\prime(k_x)\right>\left<n^\prime(k_x)\right|\alpha^\dagger_{\gamma,\sigma} \left| n(k_x)\right>}{\omega+i 0^+ - (E_{n^\prime}(k_x)-E_n(k_x))}\\\left(e^{-\beta E_n(k_x)} +e^{-\beta E_{n^\prime}(k_x)}\right),
\end{multline}
where $\alpha_{\gamma,\sigma}$ can be either $a_{s,\sigma}$ or $b_{s,\sigma}$, depending on to which sublattice  the site pointed by the $\gamma$ index belongs. Due to the two sublattice structure $s=\gamma mod(2)$, cf. Fig. \ref{lattice_pic}.
The eigenstates of the Hamiltonian in the mixed basis with a given $k_x$ are denoted by $\left|n(k_x)\right>$. 

To illustrate the properties of the low-energy excitations of the system we will consider 
\begin{equation}\label{DA_definition}
    \Delta A_\uparrow(\omega,k_x)=A_{0,\uparrow}(\omega,k_x)- A_{N/2,\uparrow}(\omega,k_x)
\end{equation}
which is the difference between the local density of states for the spin-up excitations within the same sublattice at the top edge and the middle of the ZGNR. For an edge state, expected for the non-interacting 2DTI, this quantity should give positive values for a linearly dispersive state connecting the conduction and valence bands of the bulk.

To check for gapless spin excitations we will also analyze the imaginary part of  
\begin{multline}\label{Lehman_spin}
    F^{R}_{\gamma}(\omega,k_x)=\\
    \frac{1}{Z}\sum_{n,n^\prime}\frac{\left<n(k_x)\right|S^-_{\gamma} \left|n^\prime(k_x)\right>\left<n^\prime(k_x)\right|S^+_{\gamma} \left| n(k_x)\right>}{\omega+i 0^+ - (E_{n^\prime}(k_x)-E_n(k_x))}\\\left(e^{-\beta E_n(k_x)} +e^{-\beta E_{n^\prime}(k_x)}\right),
\end{multline}
where $S^+_\gamma$ and $S^-_\gamma$ are the local spin-raising and spin-lowering operators, respectively, at site $\gamma$.
Analogously to the single-particle Greens function we introduce 
\begin{equation}\label{DF_definition}
    \Delta F(\omega,k_x)=-\frac{1}{\pi}\Im m\left\{F_{0}(\omega,k_x)- F_{N/2}(\omega,k_x)\right\}
\end{equation}
to illustrate the difference between the spin-flip excitations at the edge and in the bulk.
The numerical results discussed in this manuscript were obtained using the exact diagonalization method for a ribbon with $N=10$ sites (5 two-site unit cells). We consider only $t=1,\: t^\prime=0.2$ case at half-filling.

\section{Ribbon geometry}

For the non-interacting case, both KMM and SHM on a finite ZGNR support gapless edge modes, if the ribbon is wide enough ($s>4$) or the number of unit cells across it is odd\cite{PhysRevB.90.035116}. The latter ensures that the sublattice has a discrete translation invariance in the $y$ direction and no effective inter-edge coupling is generated. The former ensures the overlap between the edge modes is not strong enough to trigger hybridization.
To simultaneously show the localized nature of these edge modes and their direction we use $\Delta A_\uparrow(\omega,k_x)$, defined in Eq. \ref{DA_definition}. Thanks to the symmetry, we do not need to display the same spectrum for SHM, for which the difference would only be visible in the $\Delta A_\downarrow(\omega,k_x)$, due to lack of the time-reversal symmetry. 

\begin{figure}
    \centering
    \includegraphics[width=0.97\linewidth]{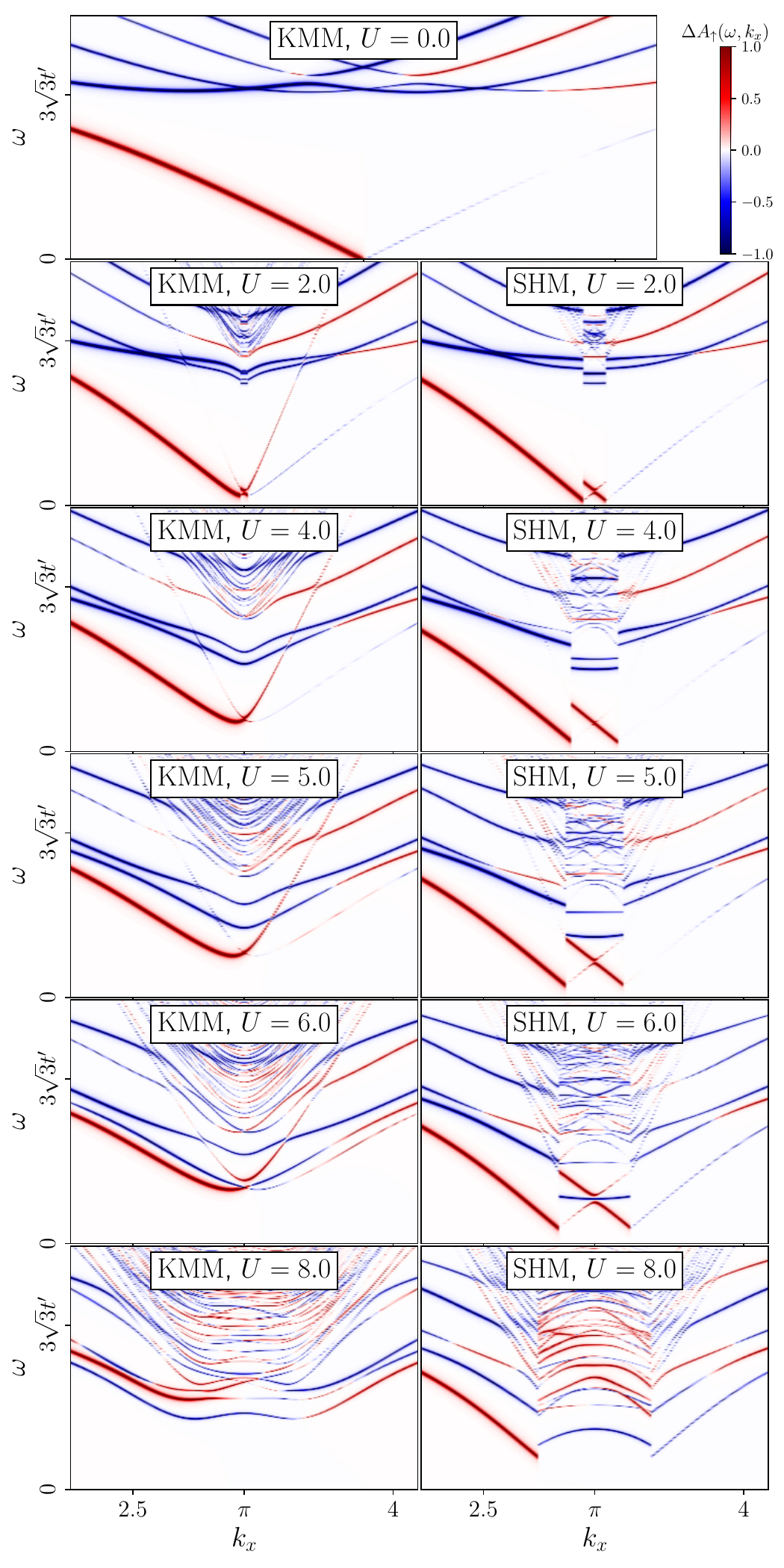}
    \caption{ A side-by-side comparison of the difference between the edge and bulk local density of states ($\Delta A_\uparrow(\omega,k_x)=A_{0,\uparrow}(\omega,k_x)- A_{N/2,\uparrow}(\omega,k_x)$) of KMM (left) and SHM (right) for various interaction strengths varying form the non-interacting (top) to the strong-coupling limit (bottom). The two models share the same non-interacting spectra, hence only one panel in the top row.}
    \label{combined_dispersions}
\end{figure}

As expected, $\Delta A_\uparrow(\omega,k_x)$ shows a linearly dispersing and left-moving edge-mode crossing the Fermi level at $k_x=\pi$ \cite{PhysRevLett.95.226801}. There is a small negative contribution (blue) in the low-energy part for $k_x>\pi$, which reflects weak tunneling of the edge state from the other end of the ribbon. An effect stemming from a relatively small ribbon width considered here. Upon increasing the system width the blue band will eventually fade away\cite{PhysRevB.90.035116}. The left-moving edge-state has exclusively positive spectral weight in the full range of $k_x$, highlighting its localized nature. The bulk bands ($\omega>3\sqrt{3}t^\prime$ ), on the other hand, change their colors revealing their spread across the whole ribbon. Thus, $\Delta A_\uparrow(\omega,k_x)$ will later be a useful tool for gauging the penetration depth of edge modes upon changing $U$.
We first focus on general changes to the spectrum as the system evolves toward the strong coupling limit and return to the onset of $U$ later.

From the second row onward, the panels in two columns of Fig. \ref{combined_dispersions} show $\Delta A_\uparrow(\omega,k_x)$ of KMM (left) and SHM (right) for $U\neq 0$. We remind here, that the true interaction strength of the analogous periodic model would be $U/4$, due to a two sub-lattice geometry. Thus $U=2$ generates terms with strength smaller than half of the non-interacting bulk gap ($3\sqrt{3}t^\prime$). Already at this $U$, both models develop a charge gap in the spectrum. As a result of interaction, the initially directional edge mode gains a branch of excitations propagating in the opposite direction. For the KMM it goes all the way across the bulk gap of the noninteracting model and there hybridizes with the bulk bands; cf. the change in color for $k_x>\pi$ and $\omega\approx3\sqrt{3}t^\prime$. This newly formed counter-propagating branch has a larger velocity. The same happens to the edge mode on the other and of the ribbon, as echoed by the emergence of a similar, but mirror reflected with respect to the $k_x=\pi$ line, low energy branch with a much smaller spectral weight located in the bulk. In the case of the SHM new bands also form, but they exist only in narrow segments of the $k_x$ range centered around the $k_x=\pi$ and conserve the linear dispersion of the non-interacting model. The existence of isolated segments of bands is allowed by the diagonal-in-$k_x$ form of the H-K interaction. For $U=2$, the bulk bands in the KMM spectra also show a discontinuity around the $k_x=\pi$ point, which at larger $U$ vanishes as will be explained later. Comparing the two spectra for the same interaction strengths, one can see that the gap in the SHM is narrower than for the KMM. The origin of this narrower gap and the shape of the bands at the gap edge will be addressed in the next section, where low-energy effective models will be introduced.

Following the evolution of $\Delta A{\uparrow}(\omega,k_x)$ with $U$ one can see, that while the gap between the edge modes increases the gap between the bulk (blue) bands becomes narrower. For $U<6$ the spectrum can be still decomposed into low-energy edge-localized bands and higher-energy bulk contributions. Between $U=5$ and $U=6$, the crossing point of the counter-propagating edge-modes, i.e. the charge gap at $k_x=\pi$, is intersected by the bulk bands. From that point on all bands have a mixed bulk-edge nature. For the discontinuous bands of SHM, a similar behavior is observed. For $U=5$ one can see the two parallel branches of the edge modes and a bulk band edge. At $U=6$ the two edge modes are crossed by the blue band of the bulk.

Also, between $U=5$ and $U=6$ the overall charge gap of the bulk bands, blue bands in Fig.\ref{combined_dispersions}, reaches the minimum value and later only grows with $U$. The initial narrowing followed by an increase of the bulk gap with $U$ around the $k_x=\pi$ point is a feature of the periodic version of these models, as will be discussed later when the projected spectra are discussed. 

Despite having vastly different spectra, both models display the same scheme of the decay of localized edge states. Namely, the lowest energy band changes from localized at the edge to mixed edge-bulk as the bulk gap and edge gap become equal at $U_c$. For $U>U_c$ the charge gap of the system only grows and there are no more localized modes at any energy. A scenario that is vastly different from the one realized in Hubbard\cite{PhysRevB.98.045133} of Falicov-Kimball model\cite{PhysRevB.106.054209}, where a charge gap is closed before the edge modes vanish.

To understand the different behaviors of the low-energy bands observed in KMM and SHM it is helpful to consider their respective effective models. We postulate them by neglecting the bulk and focusing only on the two (strongly) localized edge modes with linear dispersion coupled solely by the H-K interaction. Effectively reducing the system described by a Hamiltonian in mixed representation to a problem of two sites with $k_x$-dependent local potentials directly coupled by various terms originating from the center-of-mass conserving interaction. We neglect any (effective) inter-edge tunneling through the bulk, as the edge modes show exponential decay with the ribbon width while the inter-edge direct coupling from H-K interaction has power law dependence. Additionally, from the numerical results, one can see that there is a clear separation of energy scales between the edge and the bulk excitations. In reality, this approach is similar to the one used in the early works on the impact of correlations on the edge modes\cite{PhysRevB.79.235321, PhysRevLett.101.196804}, where a quasi-2D problem of 2DTI with edges is reduced to an effective 1D model and serves as a starting point for the bosonization. Here we have a diagonal in $k_x$ interaction, which bypasses the need for bosonization and allows for simpler treatment. A non-trivial aspect of both SHM and KMM on a ZGNR, captured by our effective models is, that the symmetry-protected existence of counter-propagating modes in real space introduces coupling between opposite quasi-momenta. A feature not usually associated with the H-K interaction. 

\section{Effective models}
The only difference between the effective models of KMM and SHM is in the location of the directional spin-dependent edge modes. For the KMM the effective kinetic part of the Hamiltonian of the edge modes is given by 
\begin{multline}\label{Eff_KMM}
    \mathcal{H}^{eff}_{KMM}(k)=-k \left(n_{\{k,0\},\uparrow} +n_{\{k,N-1\},\downarrow}\right) \\ + k \left(n_{\{k,N-1\},\uparrow} +n_{\{k,0\},\downarrow}\right)
\end{multline}
and for the spinful Haldane model, an analogous effective Hamiltonian is given by
\begin{multline}\label{Eff_SHM}
    \mathcal{H}^{eff}_{SHM}(k)=k \left(n_{\{k,0\},\uparrow} +n_{\{k,0\},\downarrow}\right)\\ + k \left(n_{\{k,N-1\},\uparrow} +n_{\{k,N-1\},\downarrow}\right).
\end{multline}
In both cases $n_{\{k,i\},\sigma}$ represents the number operator of fermions with crystal momentum $\pm k=\pm k_x+\pi$ at site $i$ and with spin $\sigma$. 
The effective SHM has $k$ and $-k$ states at either end independently of the spin projection. The effective KMM has to conserve the time-reversal symmetry and the $k$ and $-k$ states for the $\sigma$ spin channel are distributed on opposite ends to the $-\sigma$ spin channel. 

The edge modes in the two effective models are coupled by all quartic terms allowed by the H-K interaction. These are  
\begin{multline}\label{dimer_H_HK}
\mathcal{H}_{HK}^{eff}(k)=\frac{U}{4}\sum_{\sigma} \left[\sum_{\alpha\in\{0,N-1\}}\right.\\
\left(n_{\{k,\alpha\},\sigma}n_{\{k,\alpha\},\bar{\sigma}}- n_{\{k,\alpha\},\sigma}+ n_{\{k,\alpha\},\sigma}n_{\{k,\bar{\alpha}\},\bar{\sigma}}\right)+\\
\left.+\frac{1}{2} c^\dagger_{\{k,N-1\},\sigma}c_{\{k,0\},\sigma}c^\dagger_{\{k,0\},\bar{\sigma}}c_{\{k,N-1\},\bar{\sigma}}\right]
\end{multline}
The bar over $\sigma$ and $\alpha$ indicates the other value from the two elements set. The additional halving of the interaction strength, compared to the dimer model\cite{PhysRevB.109.165129}, comes from the two-atom unit cell of the graphene-like lattice considered here. Terms in Eq. (\ref{dimer_H_HK}) are grouped into density-density (second row) terms and spin-exchange (last row). It should be stressed that the former are not typical density-density interactions, where the potential energy is affected by the total charge distribution. Characteristically for the H-K interaction only the accumulation of opposite spins can increase the potential energy, irrespective of the distance. This fact lies at the heart of the ferromagnetic instability in the system with this kind of interaction.
As SHM is a direct sum of Haldane models, for each spin channel, the edge states will favor the doubly occupied edge sites in the ground state. The KMM, on the other hand, will favor a more evenly distributed charge. This has strong implications from the H-K interaction standpoint. The favoring of doubly occupied edge sites in the effective SHM (at $U=0$) means that the interaction will have a smaller overall effect on the system. From all the terms in the Hamiltonian (\ref{dimer_H_HK}) only the first density-density term and the others will be inactive.  
The uniform distribution of charge between the edge modes in KMM does not change the overall strength of the density-density terms, now the second one is active, but it also activates the spin-exchange term. As a result, one should expect stronger interaction effects in the effective KMM than in the effective SHM.
In addition, the doubly occupied sites belong to the low-spin configuration, while $U$ in a H-K dimer is known to favor high-spin solutions\cite{PhysRevB.109.165129}. 
Thus, one can expect a sharp change in the ground state upon increasing $U$ in the SHM and not in KMM. The in-depth look at the structure of the Hamiltonians in subspaces with a fixed number of particles can be found in Appendix \ref{Effcive_model_app}.

\begin{figure}
    \centering
    \includegraphics[width=\linewidth]{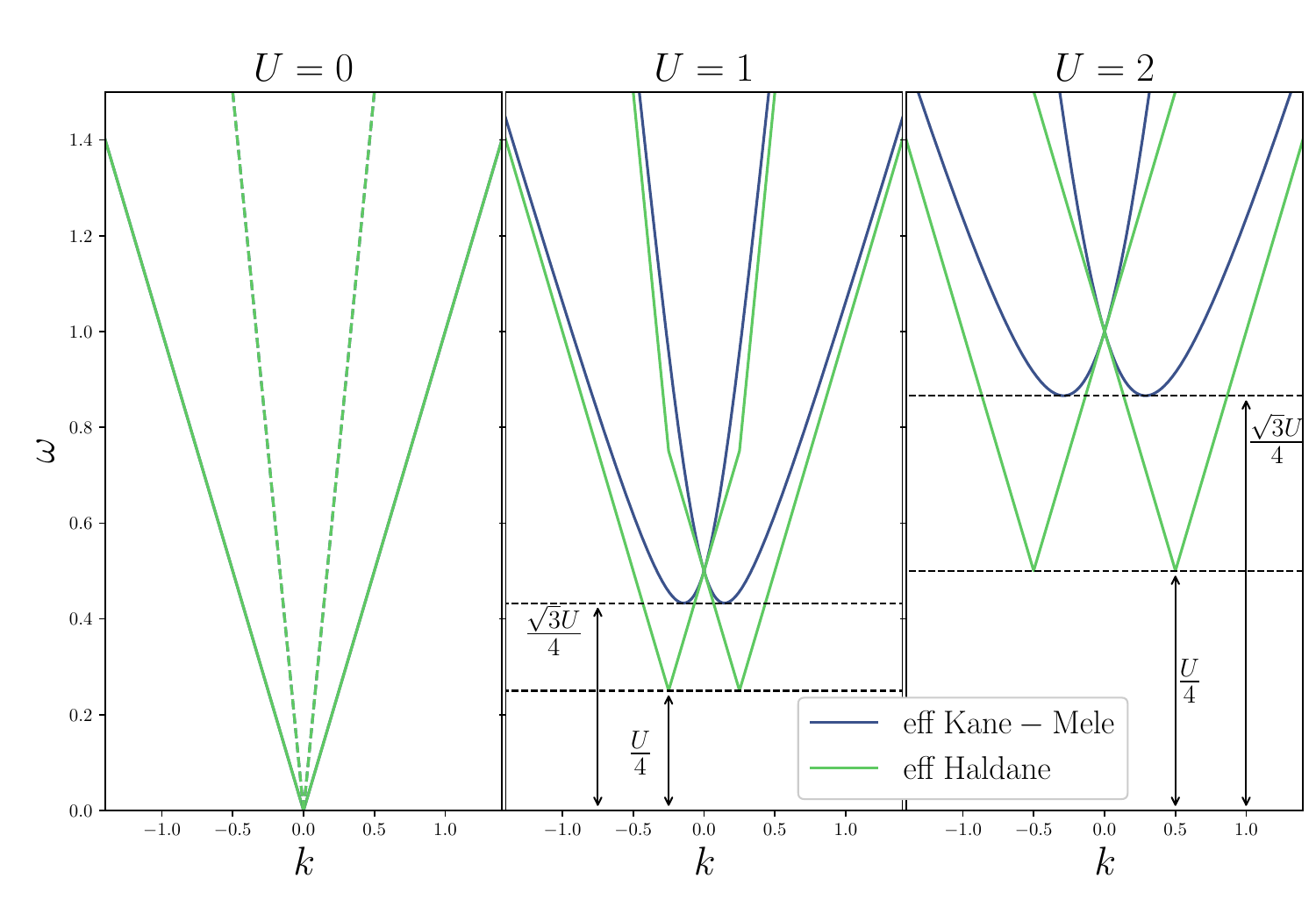}
    \caption{Location of the single particle Greens function poles of the effective model for KMM (blue) and SHM (green) edge states coupled by the H-K interaction for the three interaction regimes. In the non-interacting limit (left panel) the $3k$ mode is plotted with dashed lines as it carries zero spectral weight.}
    \label{Effective_model:spectrum}
\end{figure}

Here we will focus only on the position of the poles of the single particle Green's function of the two models, shown in Fig. \ref{Effective_model:spectrum}, and compare it to the results of the respective ZGNR models. At $U=0$ both effective models have the same spectrum consisting of technically two branches of excitations. The textbook topological edge state is one with the dispersion proportional to $\pm k$. The other behaves as $\pm 3k$ and carries no spectral weight (hence dashed lines in the left panel of Fig. \ref{Effective_model:spectrum}). A non-zero value of this single-particle transition element would mean scattering between the edge states, which is not allowed.

In both cases, adding interactions immediately leads to an avoided crossing between the edge states and opening the gap at the Fermi level. In the electron part of the single-particle spectrum ($\omega>0$), shown in Fig. \ref{Effective_model:spectrum}, the "V"-shaped crossing of two counter-propagating edge states is now split symmetrically around $k=0$ and pushed away from the Fermi level. The previously invisible excitation branch now accumulates spectral weight, cf. no colored dashed lines in two right-most panels, and asymptotically approaches the non-interacting dispersion ($\pm k$). In addition, the effective KMM displays a noticeable smoothening of the dispersion at the gap edge while the original cusp remains in the effective SHM. The qualitative agreement with the low-energy results from the respective full model calculations in Fig. \ref{combined_dispersions} suggests that the physics encapsulated in the proposed effective models is sufficient to explain the main differences between the interacting SHM and KMM.

From the analysis of the respective Hamiltonians in a two-electron subspace, cf. Appendix \ref{Effcive_model_app}, one can see that the two seemingly similar models respond very differently to the onset of COM-conserving H-K interaction. 
In the ($U,k$)-parameter space, the effective SHM has whole regions where the ground state is either in the $S=1$ or $S=0$ subspace, where $S$ is the total spin. The high-spin solution lacks any $k$ dependence since it is built out of pairs of electrons with the same spin projection, which are counter-propagating. The ground state in the $S=0$ subspace is adiabatically connected to the non-interacting solution and reflects the immanent favoring of the doubly occupied edges of the SHM. 
The former is the ground state for $|k|< U/4$ and the latter outside that range. The lack of $k$ dependence in the high-spin ground state energy means local single-electron excitation will carry its linear dispersion. But adding/removing an electron costs $U/4$ energy that binds the high-spin solution through the spin-exchange term and gives rise to the gap. The effective model shows that the separation between the cusps grows linearly with $U$, cf. Appendix \ref{Effcive_model_app}, which is also the case for the full SHM spectrum.

The effective KMM also has a ground-state degeneracy region in the ($U,k$)-space, but it is restricted to the $k=0$ line only, and nowhere is the ground-state limited to the $S=1$ subspace. Outside the $k=0$ line, the non-degenerate interacting ground state is adiabatically connected to the non-interacting one. As mentioned before, the time-reversal symmetry of the effective KMM promotes equal distribution of electrons with opposite spins on opposite edges. This activates terms in the H-K interaction, that allow for inter-edge scattering, which in turn is giving mass to the edge modes. For details see Appendix \ref{Effcive_model_app}. The linear in $U$ splitting of the band minima from the effective model agrees with the full model calculation, similar to the SHM case.

From the effective models, one can also recover the narrower charge gap in SHM than in the KMM. The exact ratio of $\sqrt{3}$ of the gaps is not reproduced in the full model calculations. This is possibly due to hybridization effects that generate avoided crossing of the edge modes not captured by the effective models, compare spectra of Fig. \ref{combined_dispersions} at $k_x=\pi$ and Fig. \ref{Effective_model:spectrum} at $k=0$.
The last property of the full model reproduced by its effective analog is that the crossing of the two branches of the low-energy modes is the same in both cases, despite different gap widths and splittings of the bands' minima. This property holds as long as the edge modes are localized ($U<3$). Due to the lack of the bulk, the effective model fails to recover the anti-crossing of the edge modes. Most likely, the long-range nature of the H-K interaction generates additional effective coupling between the edges through the bulk sites, which is responsible for this effect.

The fact that in both cases the study of the effective models reproduces qualitatively the numerical results and gives a charge gap $\propto U/N$ suggests, that for any finite width of the ribbon, the gapless edge modes should be unstable and the system will be an insulator with a charge gap. This result is consistent with other reports on the stability of the edge modes in the presence of either long-range interaction\cite{PhysRevLett.101.196804,PhysRevB.73.045322} or local interactions but with the addition of an effective coupling between the edges\cite{PhysRevB.90.035116}. These reports point out, using bosonization, that while the charge gap could be open because of effective coupling between the edges, either through a single- or two-particle process, there is a possibility of sustaining spin-conducting edge modes. 

\section{Spin edge modes}

\begin{figure}
    \centering
    \includegraphics[width=0.98\linewidth]{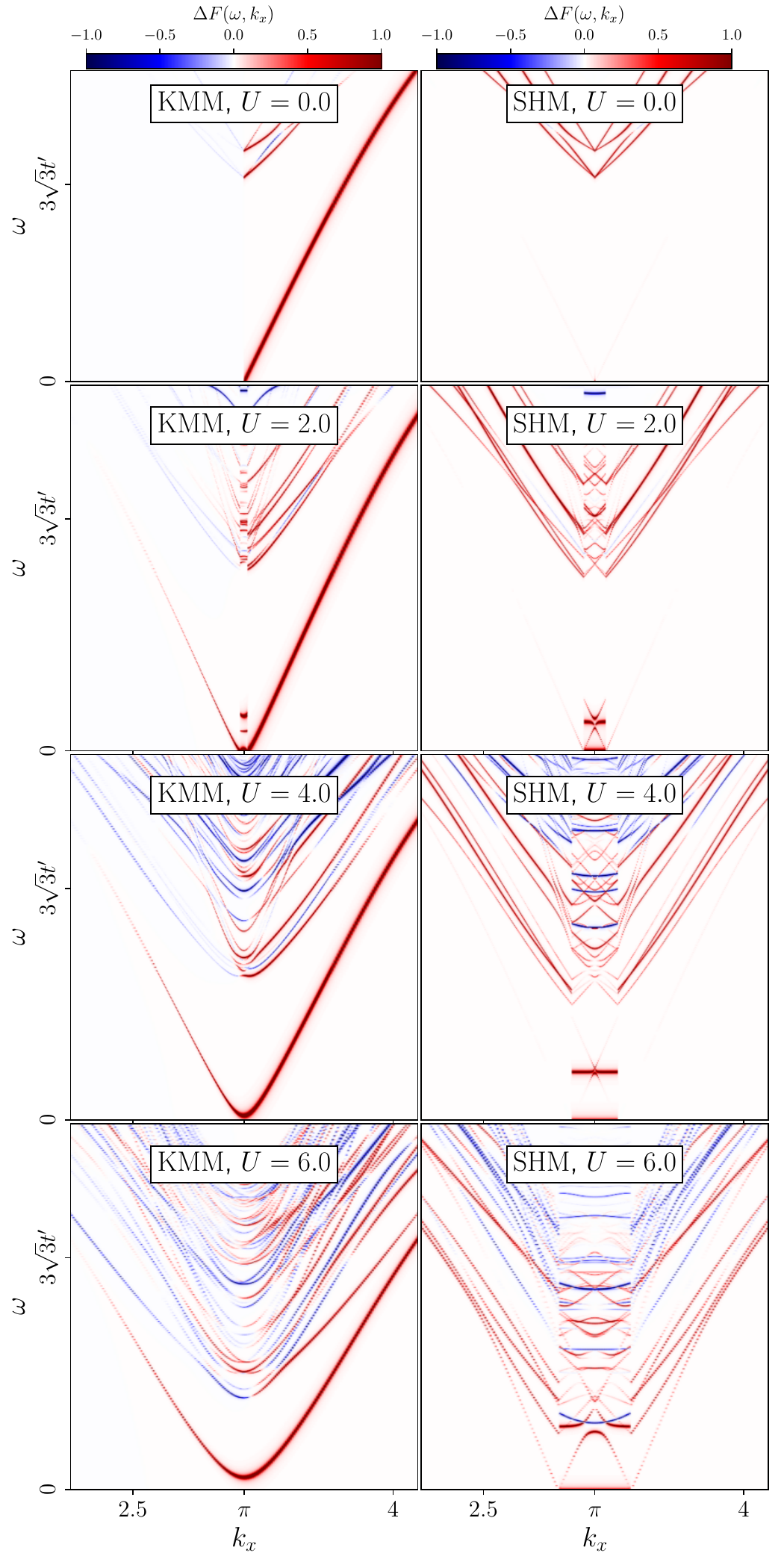}
    \caption{Comparison of the difference between the edge and bulk spin spin-excitation spectrum between the KMM (left) and SHM (right) for various interaction strengths $U$. The red color indicates the edge-localized mode while bulk modes are in blue. }\label{spin_spectrum}
\end{figure}

To check for the existence of gapless spin excitations in the system we analyze the spectrum of the spin analog of the single-particle Green's function defined in Eq. (\ref{Lehman_spin}). To see the penetration depth of these spin excitations we will look again at the difference between the local spectrum at the edge and in the bulk. The results are displayed in Fig.  \ref{spin_spectrum}, with KMM results in the left column and SHM in the right one. 

This time the spectra of the non-interacting KMM and SHM are different. The former has a well-resolved momentum-locking of the spin-excitations, as indicated by the asymmetric, $k_x>\pi$ vs $k_x<\pi$, distribution of the spectral weight of the edge mode. Because in $\Delta F(\omega,k_x)$ we consider $F^R_{\gamma=0}(\omega,k_x)$ defined in Eq. (\ref{Lehman_spin}), 
that describes transition caused by a local spin-flip, the spin-momentum locking gives a branch with only positive velocity. 
The SHM spin spectrum has a symmetric distribution of the spectra, with the lowest energy branch barely visible in the top panel of Fig. \ref{spin_spectrum}. This weak branch is only present due to inter-sublattice hybridization effects. From a simplified picture based on the effective model, a non-interacting SHM should have local spin-flip excitation blocked by the Pauli principle. For this model, an inter-edge spin-flip excitation is preferred, but a local object like $F^R_{0}(\omega,k_x)$ is not capable of capturing it.
Nonetheless, both models in the considered geometry have gapless local spin excitations at $k_x=\pi$. 
At $U=2$ (second row of Fig.\ref{spin_spectrum}) both models still have gapless spin modes, while having a well-developed charge gap for the same interaction strength, cf. second row of Fig. \ref{combined_dispersions}. That is in line with predictions from the renormalization analysis done for other, more traditional, types of interaction.
In the case of KMM, the dispersion touches zero in only two $k_x$ points located symmetrically around the $k_x=\pi$, for which the charge gap was narrowest. As the interaction strength increases to $U=4$, this band becomes parabolic and the spin gap is opened. In SHM the situation is different, in the low-energy sector the spin excitations are dispersionless around the $k_x=\pi$ point and acquire significantly more spectral weight compared to the $U=0$ case. Increasing $U$  widens the flat band segment, which remains pinned to the Fermi level. 
 
This flat spin-excitation branch echoes the ground-state degeneracy of the high-spin solution found between the kinks in the single-particle dispersion of SHM, cf. Fig. \ref{combined_dispersions}. The particle number conservation in a spin-flip process paired with the degeneracy of the ground state results in a zero-energy branch of the spin excitations. For the effective SHM, the non-zero spectral weight of the zero-energy branch comes from the transitions between two $S=1$ states, $S_z=-1 \rightarrow S_z=0$. Outside the region with a high-spin ground state, the lack of degeneracy means there is no more pinning of the spin excitations to the Fermi level and they behave as in the non-interacting limit.

In the case of weakly interacting KMM the effective model proves to be insufficient to explain the gapless spin-excitations. The non-degenerate ground state of the effective model is a mixture of two $S_z=0$ and the expectation value of the spin-flip operator has to vanish. Thus gapless spin excitations are not allowed. This inconsistency is observed only for weak interactions $U<4$. For $U\leq 4$ the direct gap at $k_x=\pi$ matches qualitatively the predictions of the effective model i.e. parabolic band with a gap centered at $k=0$. The fact that this lack of agreement between the full and effective models is observed only for the KMM and in the weak-$U$ regime, points towards neglecting the real-space location of the spinful edge modes as its origin. In the SHM they are located at the same sites for either spin channel, so the two-sublattice structure of the lattice is not important and the effective model description works well. For the KMM, they reside on opposite sublattices. When $U<4$ the inter-sublattice $t$-contributions are of similar strength to $U$-terms and the internal dynamics of a unit cell at each edge becomes relevant. Approximating it by the dimer physics one would expect  
the formation of local singlets at the edges for small $U$ \cite{PhysRevB.109.165129}. This would explain a non-zero expectation value of a spin-flip. Following the local dimer analogy, the reduction of the singlet contribution to the ground state of a dimer, in favor of an anti-bonding state of doubly occupied sites, explains the decreasing importance of the sublattice structure of the underlying model for larger $U$.

In both cases of the 2DTI, the gapless spin excitations are well localized at the edges of the ribbon, as can be inferred from the high positive value of $\Delta F_\uparrow(\omega,k_x)$ in Fig. \ref{spin_spectrum}. The bulk bands are only visible at much higher energies.

\section{Periodic lattice}

To finish the analysis of the edge states we turn to the problem of the {\it bulk-boundary correspondence} in models with H-K interaction. Despite KMM being often used to analyze the competition between the topology and correlation, previous studies exploring the impact of H-K interaction on topology have not explicitly considered the full form of the original interaction. Most noticeably the non-density-density inter-sublattice interaction terms (spin-exchange) have not been explicitly analyzed. Previous studies either included only terms diagonal in the sublattice basis (orbital H-K) or only of the density-density form (band H-K). From the point of view of the H-K interaction, these spin-exchange terms have to be included, as they also conserve the center-of-mass property, which is defining principle of the HK model. Hence, first, we will focus on the role played by these terms in the fate of the topological phase transition in both models and then move to the analysis of the {\it bulk-boundary correspondence}. 

The demise of the non-trivial topological phase in the presence of H-K-like interaction originates from the closing of the gap in the many-body spectrum and not in the single particle one\cite{PhysRevB.108.035121}. As the change of the ground state spreads over the BZ with increasing $U$ the topological invariant continuously goes to zero\cite{PhysRevB.110.075105} despite the system being an insulator all the time. This behavior is in contrast to the non-interacting or local in space interaction, where the TPT is signaled by closing of the charge gap in the single particle excitation spectrum. For systems with H-K-like interaction, this spectrum can only display kinks or discontinuities across the TPT, while the charge gap remains open\cite{PhysRevB.108.035121,PhysRevB.108.195145,PhysRevB.110.075105}. Their origin is inherently connected to the locality in the $k$-space of the Hamiltonian, which makes the solutions for each momentum $k$ independent. Because the breakdown of the topology comes from the change in the ground state (GS) we will use the GS degeneracy within the BZ zone as a proxy for topology.

\begin{figure}
    \centering
    \vspace{0.3cm}
    \includegraphics[width=0.95\linewidth]{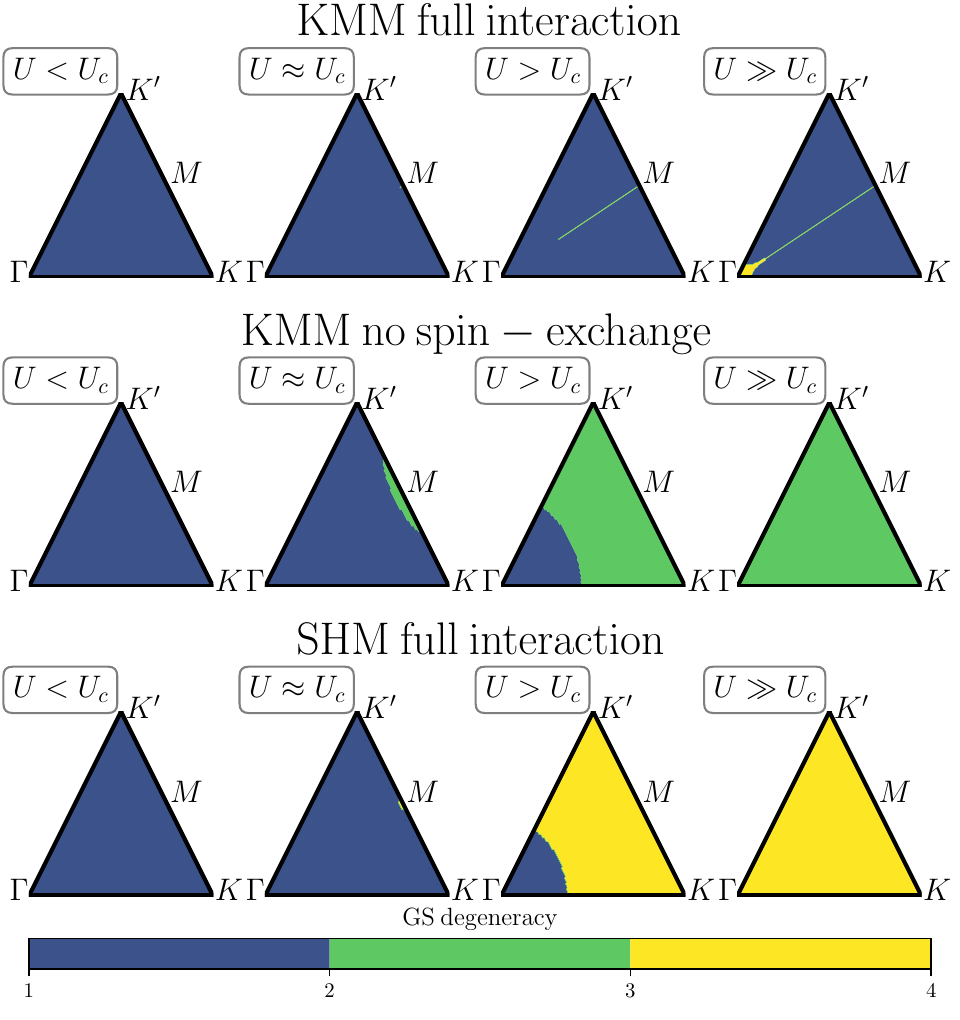}
    \caption{Comparison of the evolution of the ground-state degeneracy inside the BZ of the KMM with increasing $U$ with the full interaction (top) and the interaction without the inter-sub-lattice coupling terms(middle). In the bottom row are the corresponding results for the SHM with full interaction.}\label{GS_deg}
\end{figure}

Figure \ref {GS_deg} shows how the ground state degeneracy inside the one-sixth of the Brillouin zone (BZ) changes with interaction strengths for the half-filled KMM with spin-exchange terms (top), KMM without the spin-exchange (middle) and for the SHM with spin-exchange. We do not display the exact values of $U$, as they depend on the specifics of the models, and use $U_c$ as the critical interaction strength at which the GS degeneracy appears. For the full H-K interaction and $t^\prime=0.2$, both KMM and SHM have $U_c\approx 5.66$, while for the KMM and H-K-like interaction (without the spin-exchange) $U_c=4*2.078\approx 8.3$\cite{PhysRevB.110.075105}. The rescaling factor of four comes from the square of the number of sub-lattices, not used explicitly in the other work\cite{PhysRevB.110.075105}. From Fig. \ref{GS_deg} one can see that in all cases the breakdown of the integer topological invariant starts at the $M$-point and later spreads across the full BZ as $U$ increases. The fact that it starts at the $M$-point is accidental. For the two bottom cases, the degeneracy appearing first at the $M$-point for $t^\prime\ge 0.2$ coincides with the single-particle and many-body spectral gap being narrowest there. For $t^\prime<0.2$ the gap is minimal at $K/K^\prime$ points and that is where the ground-state degeneracy starts to spread from.

For the KMM with all interaction terms, the situation is different. The ground-state degeneracy starts at the $M$-point independent of $t^\prime$. It first spreads only along the $M\rightarrow\Gamma$ path. Eventually, increasing $U$ makes the degeneracy region spread around the $\Gamma$ point. At $K/K^\prime$ point, the GS remains non-degenerate irrespective of $U$. This means that the spin-Chern number is never fully suppressed by the interaction. To explain these behaviors it is best to look at the Hamiltonian matrix of each model in the two-electron subspace. Using the notation from Ref. \onlinecite{PhysRevB.108.165136}, the KMM with full H-K interaction is given by 
\begin{widetext}
\begin{equation}\label{PBC_H2}
   \mathcal{H}_{KMM,2}(\vec{k})= \left[
    \begin{array}{cccccc}
        -\frac{U}{2} & 0 & 0 & 0 & 0 & 0  \\
        0 & -\frac{U}{4} & -tg^*(\vec{k}) & tg^*(\vec{k}) & 0 & 0  \\
        0 & -tg(\vec{k}) &-\frac{U}{4}-4t^\prime g_{1}(\vec{k}) & -\frac{U}{4} & -tg^{*} (\vec{k}) & 0 \\
        0 & tg(\vec{k}) &-\frac{U}{4} & -\frac{U}{4} +4t^\prime g_1(\vec{k}) & tg^{*}(\vec{k}) & 0 \\
        0 &  & -tg(\vec{k}) & tg(\vec{k}) & -\frac{U}{4} & 0  \\
        0 & 0 & 0 & 0 & 0 & -\frac{U}{2} \\ 
    \end{array}
    \right],     
\end{equation}
\end{widetext}
where the $g(\vec{k})=\sum_{j} e^{i\vec{k}\cdot\vec{a}_j}$ with $\vec{a}_j$ vectors connecting all the nearest neighbour sites, $g_{1}(\vec{k})=\sum_j\sin(\vec{k}\cdot\vec{b}_j)$ with $\vec{b}_j$ vectors connecting all the next-nearest neighbour sites. The $\mathcal{H}_{KMM,2}(\vec{k})$ matrix is written using following basis states:
\begin{tasks}[style=enumerate](3)
    \task $a^\dagger_\uparrow b^\dagger_\uparrow\left| 0\right \rangle$
    \task $a^\dagger_\uparrow a^\dagger_\downarrow\left| 0\right \rangle$
    \task $a^\dagger_\downarrow b^\dagger_\uparrow\left| 0\right \rangle$
    \task $a^\dagger_\uparrow\ b^\dagger_\downarrow\left| 0\right \rangle$
    \task $b^\dagger_\uparrow b^\dagger_\downarrow\left| 0\right \rangle$
    \task $a^\dagger_\downarrow b^\dagger_\downarrow\left| 0\right \rangle$
\end{tasks}
The spin-exchange term introduces a $-\frac{U}{4}$ coupling between the third and fourth states in the center of the matrix (\ref{PBC_H2}). At $K/K^\prime$ points $g(\vec{k})=0$, $g_{1}(\vec{k})=\pm 3\frac{\sqrt{3}}{2}$ and the only off-diagonal terms are the spin-exchange ones. Without them, there is a critical interaction $U_c =24\sqrt{3}t^\prime$ at which the initially non-degenerate ground state has the same energy as the first and the last basis state, hence for larger $U$ the ground state becomes doubly degenerate, as shown in Fig. \ref{GS_deg}. The presence of the spin exchange ensures that the ground state remains in the subspace spanned by the third and fourth states and remains non-degenerate. 
This is not the case for the SHM, for which the $\pm 4t^\prime g_1(\vec{k}) $ elements of the matrix (\ref{PBC_H2}) are at the diagonal (second and fifth place). As a result, the ground state jumps to the triple degenerate subspace already for $U\ge 12\sqrt{3}t^\prime$, cf. bottom row of Fig. \ref{GS_deg}. 

From the form of the Hamiltonian (\ref{PBC_H2}) we get that if both models have ground-state degeneracy spreading from the $M$-point then they have the same $U_c$. This is because at the $M$-point the kinetic part of the Hamiltonian of both models reduces to the one of graphene due to vanishing of $g_{1}(\vec{k}=M)$. Nowhere else in the BZ this takes place. Moreover one can calculate that this matching of $U_c$ happens when $t^\prime > t/3\sqrt{3}$.

\section{Bulk-boundary correspondence}

After establishing the properties of the periodic models' ground state we return to the question of the {\it bulk-boundary correspondence} in systems with H-K interaction. As discussed in the previous Section for $t^\prime=0.2$ the two periodic models of 2DTI have the same $U_c\approx 5.66$. This interaction strength is within the range of $U$ for which the numerical results for the ribbon geometry showed the onset of hybridization between the edge and the bulk bands. Due to the limitation of numerical calculations, it is not possible to confirm it with absolute certainty. Still, the presented results strongly indicate that the robustness against the interaction of the topological phase of a 2DTI correlates with the existence of localized modes at the edges of a sample. The localization of a mode is understood as the existence of a band with the majority of its spectral weight coming from excitations at a certain site across the ribbon for all $k_x$. Contrary to the usual edge states, these localized modes are charge-gapped. In the case of time-reversal lacking SHM, the system can have zero energy spin excitation that reflects the degeneracy of the high-spin ground state favored by the H-K interaction in finite geometries. The time-reversal symmetry of KMM resists the formation of a high-spin (degenerate) state. Only in the limit of weak-$U$ the intra-sublattice dynamics allow for gapless spin excitations, that turn gapped once the non-local H-K interaction terms start to dominate.

\section{Conclusions}
In this work, we discussed the spectral properties of two models of 2DTI, the spinful Haldane and Kane-Mele models, in the presence of COM-conserving H-K interaction on a ribbon geometry. These two models, which differ by topological invariants in the non-interacting limit, host gapless edge states. Using the exact diagonalization method we show, that in a ribbon with 10 sites across the long-range nature of H-K interaction leads to the opening of the charge gap. Below a certain interaction strength, the gapped edge-modes remain localized, but depending on the symmetry of the underlying 2DTI these modes can be parabolic-massive (KMM) or linear-massless (SHM). Using an effective model of the two, we explain the observed differences and argue that the same behavior would be observed in much larger systems and for any finite $U$. We also look at the spin excitations and confirm they can be gapless. For the KMM only for weak interactions and for SHM for any $U$. Lastly, we address the issue of the {\it bulk-boundary correspondence} for the interaction that, in principle, changes with the boundary conditions. We argue, that it holds to a certain degree. Instead of observing the disappearance of the edge modes upon the gap reopening at the TPT, we report the onset hybridization between the edge and bulk modes as bulk loses its integer topological invariant.

This research was partially supported by the “MagTop” project (FENG.02.01-IP.05-0028/23) carried out within the “International Research Agendas” programme of the Foundation for Polish Science co-financed by the European Union under the European Funds for Smart Economy 2021-2027 (FENG) and by Narodowe Centrum Nauki (NCN, National Science Centre, Poland) Project No. 2019/34/E/ST3/00404. We acknowledge the access to the computing facilities of the Poznan Supercomputing and Networking Center Grant No. 609.

\bibliographystyle{apsrev4-2}
\bibliography{biblio_edege_states}

\newpage
\appendix
\section{Hatsugai-Kohmoto in mixed basis representation}\label{HK_mixed_basis_form}
For a honeycomb lattice, like the KMM or SHM, the primitive shifts are given by
$$
\hat{x}=(a,0 )\:, \:\hat{y}=(\frac{a}{2},\frac{\sqrt{3}a}{2})
$$
and positions are given by 
$$
j_i=x_i\hat{x} +y_i\hat{y},
$$
where $x_i$ and $y_i$ are integers.
In addition, the unit cell has two atoms, which we will denote using Latin letters $a,b$. Because the primitive shift are not orthogonal a little more attention has to be paid to the indices, which will be discussed later. First one has to exploit the conservation of center-of-mass within the two-atom unit cell. The Hatsugai-Kohmoto Hamiltonian in the two-atom unit cell has three pairs of terms, which can be written explicitly as
$$
\mathcal{H}_{H-K}=\frac{U}{(2N)^2}\sum_{j_1,j_2,j_3,j_4}\delta_{j_1+j_3-j_2-j_4}\times
$$
$$
\left\{a^\dagger_{j_1,\uparrow}a_{j_2,\uparrow}a^\dagger_{j_3,\downarrow}a_{j_4,\downarrow}+b^\dagger_{j_1,\uparrow}b_{j_2,\uparrow}b^\dagger_{j_3,\downarrow}b_{j_4,\downarrow}+\right.
$$
$$
\left . a^\dagger_{j_1,\uparrow}a_{j_2,\uparrow}b^\dagger_{j_3,\downarrow}b_{j_4,\downarrow}+b^\dagger_{j_1,\uparrow}b_{j_2,\uparrow}a^\dagger_{j_3,\downarrow}a_{j_4,\downarrow} +\right.
$$
$$
\left . a^\dagger_{j_1,\uparrow}b_{j_2,\uparrow}b^\dagger_{j_3,\downarrow}a_{j_4,\downarrow}+b^\dagger_{j_1,\uparrow}a_{j_2,\uparrow}a^\dagger_{j_3,\downarrow}b_{j_4,\downarrow} \right\}
$$
The $N$ indicates the number of unit cells and not lattice sites, hence interaction strength $U$ is divided by $(2N)^2$ which is the number of all sites. In the second and third lines are terms describing (two) correlated hoppings, which do not exchange spins between the sublattices. In the second line are terms that describe both hoppings being within the same sublattice and in the third line are pairs of intra-sublattice hoppings each within a different sublattice. In both cases, the $\delta$ part alone controls the conservation of center-of-mass. In the fourth lane are the pair hoppings, allowed by the conservation of the center-of-mass principle, which effectively describes the exchange of spin between the sublattices. 
The first two types of interaction terms produced by the H-K interaction were explicitly discussed\cite{PhysRevB.108.165136,PhysRevB.110.075105}, while the "spin-exchange" was partially hidden within the introduction H-K interaction in the band basis. 
To better illustrate the transition from primitive shifts to Cartesian coordinates in the following we will focus on only one term. The transformation for the remaining terms is the same, thanks to the splitting of the terms into intra- and inter-sublattice couplings.
$$
\mathcal{H}_{H-K}=\frac{U}{4N^2}\sum_{j_1,j_2,j_3,j_4}\delta_{j_1+j_3-j_2-j_4}a^\dagger_{j_1,\uparrow}a_{j_2,\uparrow}a^\dagger_{j_3,\downarrow}a_{j_4,\downarrow}=
$$
$$
=\frac{U}{4N^2}\sum_{x_1,\ldots,x_4}\sum_{y_1,\ldots,y_4}\delta_{x_1+\frac{1}{2}y_1+x_3+\frac{1}{2}y_3,x_2+\frac{1}{2}y_2+x_4+\frac{1}{2}y_4}
$$
$$
\delta_{y_1+y_3,y_2+y_4}\cdot
$$
$$
\cdot a^\dagger_{\{x_1,y_1\},\uparrow}a_{\{x_2,y_2\},\uparrow}a^\dagger_{\{x_3,y_3\},\downarrow}a_{\{x_4,y_4\},\downarrow}
$$
The transition from the first to the second line is the splitting of multi-indices of the unit cell on the primitive shift basis to the separate conservation of COM along Cartesian $x$ and $y$ coordinates. As the Cartesian $x$ is the same as primitive shift $x$ it will set the length scale.
Now we do a Fourier transform along $x$ direction
\begin{multline*}
\mathcal{H}_{H-K}=\frac{U}{4N^2}\sum_{x_1,\ldots,x_4}\sum_{y_1,\ldots,y_4}
\\\delta_{x_1+\frac{1}{2}y_1+x_3+\frac{1}{2}y_3,x_2+\frac{1}{2}y_2+x_4+\frac{1}{2}y_4}\delta_{y_1+y_3,y_2+y_4}\cdot
\end{multline*}
$$
\cdot \sum_{k_{x,1},k_{x,2},k_{x,3},k_{x,4}} e^{ik_{x,1} x_1}e^{-ik_{x,2} x_2}e^{ik_{x,3} x_3}e^{-ik_{x,4} x_4}\cdot
$$
$$
a^\dagger_{\{k_{x,1},y_1\},\uparrow}a_{\{k_{x,2},y_2\},\uparrow}a^\dagger_{\{k_{x,3},y_3\},\downarrow}a_{\{k_{x,4},y_4\},\downarrow}=
$$
$$
=\frac{U}{4N^2}\sum_{y_1,\ldots,y_4}\delta_{y_1+y_3,y_2+y_4}\sum_{x_1,\ldots,x_4}\frac{1}{N}\sum_{k_x}\cdot
$$
$$
\cdot e^{-i(x_1+x_3-x_2-x_4)k}e^{-\frac{i}{2}(y_1+y_3-y_2-y_4)k_x}\cdot
$$
$$
\cdot \frac{1}{N^2}\sum_{k_{x,1},k_{x,2},k_{x,3},k_{x,4}} e^{i k_{x,1} x_1}e^{-i k_{x,2} x_2}e^{i k_{x,3} x_3}e^{-i k_{x,4} x_4}\cdot
$$
$$
a^\dagger_{\{k_{x,1},y_1\},\uparrow}a_{\{k_{x,2},y_2\},\uparrow}a^\dagger_{\{k_{x,3},y_3\},\downarrow}a_{\{k_{x,4},y_4\},\downarrow}=
$$
$$
=\frac{U}{4N}\sum_{y_1,\ldots,y_4}\delta_{y_1+y_3,y_2+y_4}\cdot 
$$
$$
\sum_{k_x}e^{-\frac{i}{2}(y_1+y_3-y_2-y_4)k_x}\cdot
$$
$$
\prod_{\alpha=1}^4 \frac{1}{N}\sum_{k_{x,\alpha},x_\alpha}e^{(-1)^{\alpha+1}i(k_{x,\alpha}-k_x)x_\alpha}\cdot
$$
$$
a^\dagger_{\{k_{x,1},y_1\},\uparrow}a_{\{k_{x,2},y_2\},\uparrow}a^\dagger_{\{k_{x,3},y_3\},\downarrow}a_{\{k_{x,4},y_4\},\downarrow}=
$$

\begin{multline*}
\frac{U}{4N}\sum_{y_1,\ldots,y_4}\delta_{y_1+y_3,y_2+y_4}\sum_{k_x} e^{-\frac{i}{2}(y_1+y_3-y_2-y_4)k_x}
\\
 a^\dagger_{\{k_x,y_1\},\uparrow}a_{\{k_x,y_2\},\uparrow}a^\dagger_{\{k_x,y_3\},\downarrow}a_{\{k_x,y_4\},\downarrow}
\end{multline*}
The final form of the H-K interaction in a mixed basis is

\begin{multline*}
\mathcal{H}_{H-K}=\frac{U}{2N}\sum_{y_1,\ldots,y_4}\delta_{y_1+y_3,y_2+y_4}
\\
\sum_k e^{-\frac{i}{2}(y_1+y_3-y_2-y_4)k_x}
\\
\frac{1}{2}\left\{ a^\dagger_{\{k_x,y_1\},\uparrow}a_{\{k_x,y_2\},\uparrow}a^\dagger_{\{k_x,y_3\},\downarrow}a_{\{k,y_4\},\downarrow}+\right.
\\
b^\dagger_{\{k_x,y_1\},\uparrow}b_{\{k_x,y_2\},\uparrow}b^\dagger_{\{k_x,y_3\},\downarrow}b_{\{k_x,y_4\},\downarrow}+
\\
a^\dagger_{\{k_x,y_1\},\uparrow}a_{\{k_x,y_2\},\uparrow}b^\dagger_{\{k_x,y_3\},\downarrow}b_{\{k_x,y_4\},\downarrow}+
\\
b^\dagger_{\{k_x,y_1\},\uparrow}b_{\{k_x,y_2\},\uparrow}a^\dagger_{\{k_x,y_3\},\downarrow}a_{\{k_x,y_4\},\downarrow}+
\\
 a^\dagger_{\{k_x,y_1\},\uparrow}b_{\{k_x,y_2\},\uparrow}b^\dagger_{\{k_x,y_3\},\downarrow}a_{\{k_x,y_4\},\downarrow}+
\\
\left. b^\dagger_{\{k_x,y_1\},\uparrow}a_{\{k_x,y_2\},\uparrow}a^\dagger_{\{k_x,y_3\},\downarrow}b_{\{k,y_4\},\downarrow}\right\}
\end{multline*}
Lastly, additional phase factor $f(y_1,y_2,y_3,y_4)$ has to be added to each term in the H-K interaction, to account for a gauge transformation that brings the kinetic part of the Hamiltonians in the mixed basis into a completely real form. This transformation, introduced in Ref.\cite{PhysRevB.90.035116}, adds a phase factors $e^{ik_x/2}$ to fermionic operators that act on sites, which are on the right side of the ZGNR super-cell, depicted in Fig. \ref{lattice_pic}. The H-K interaction is not invariant under this transformation, thus the phase factor has to appear that accumulate all phase factors of a given set of four sites.


\section{Ground states of the effective models}\label{Effcive_model_app}
In this section, we provide the exact forms of the effective Hamiltonians \ref{Eff_KMM} and \ref{Eff_SHM} in the two-electron subspace. 
First, we will focus on the $\mathcal{H}^{eff}_{KMM}(k)$. Following steps as in the similar analysis of a two-site H-K model, with subspace spanned by eigenstates of the total spin operator, we obtain the Hamiltonian matrix


\begin{equation}
\left[
   \begin{array}{c c c | c c c}
    \upermat{S=0}{0 & 0 & 0} &2k &0 & 0 \\
    0 & -\frac{U}{2} & 0 &0  &0 &0\\
    0 & 0 & -\frac{U}{2} & 0 &0 &0\\
    \hline
    2k & 0 & 0 & -U &0 &0 \\
    0 & 0 & 0 & 0 & -U&0 \\
    0 & 0 & 0&\undermat{S=1}{0 & 0&-U }
    \end{array}
   \right] 
   ,
\vspace{0.5cm}
\end{equation}
where $S=0,1$ is the total spin of each state in the submatrix.
Assuming $k>0$, this matrix has a non-degenerate ground state with the energy 
\begin{equation}
    E_{GS}=-\frac{U}{2}-\sqrt{(2k)^2+\left(\frac{U}{2}\right)^2},
\end{equation}
which is formed as a linear combination of the two $S_z=0$ states, one from the spin-singlet and the other from the spin-triplet subspace. The coupling between the two $S$ subspaces originates from the fact that the location of the edge states promotes $c^\dagger_{\{k,0\},\uparrow}c^\dagger_{\{k,N-1\},\downarrow}\left|0\right>$ state, which is not an eigenstate of the total spin operator. This coupling between the spin subspaces secures the existence of a non-degenerate ground state, which is the reason behind continuous bands for all $k$ values.  

The excited states are doubly degenerate and have the energies of
\begin{equation}
    E_{EX}=\pm k -\frac{U}{2}
\end{equation}

The poles of the single-particle Greens functions have two branches, depending on the spin and location of the excited electron. Expanding the single particle excitation energy for small $k$ (up to quadratic terms in $k$) gives
\begin{equation}
    E_{EX}-E_{GS}\approx \frac{\left[k \pm \frac{U}{8}\right]^2}{2\frac{U}{8}}+\frac{7U}{16}.
\end{equation}
From this form, one can see that two branches of excitations are split by $U/4$ and acquire a mass of $U/8$. The charge gap is $7U/16$, which matches the exact result of $\sqrt{3}/4$ very well.

Similarly, the effective SHM Hamiltonian \ref{Eff_SHM} in the two-electron subspace has the following form \\
\begin{equation}
\left[
   \begin{array}{c c c | c c c}
     \upermat{S=0}{0 & 0 & 0 }&0 &0 & 0 \\
    0 & -\frac{U}{2} & -2k &0  &0 &0\\
    0 & -2k& -\frac{U}{2} & 0 &0 &0\\
    \hline
    0 & 0 & 0 & -U &0 &0 \\
    0 & 0 & 0 & 0 & -U&0 \\
    0 & 0 & 0 & \undermat{S=1}{0 & 0& -U} 
    \end{array}
   \right]
.
 \vspace{0.5cm}
\end{equation}

It consists of two disconnected blocks with different total $S$, which is responsible for the jump in the spectrum. In the case of Hamiltonian \ref{Eff_SHM} the phase transition from low-spin to high-spin state is at $k=\frac{U}{4}$. For $k$ outside that region, the ground state remains in the same $S$ subspace as for the non-interacting system. Inside that $k$ range the H-K interaction dominates and the ground state is inside the high-spin sector and is triple degenerate. At the transition point between the ground-states ($k=\frac{U}{4}$), the charge gap is 
$$
E(k=\frac{U}{4})=E_{EX}-E_{GS}=-\frac{3}{4}U+U=\frac{U}{4}
$$
and the single particle excitations conserve their linear dispersion both for $k<\frac{U}{4}$
$$
E_{EX}-E_{GS}=\pm k -\frac{U}{2}+U=\pm k+\frac{U}{2}
$$
and for $k>\frac{U}{4}$
$$
E_{EX}-E_{GS}= \pm k -\frac{U}{2}+2 k+\frac{U}{2}=(2 \pm 1)k$$
The former shows that there exists a gap and that the bands of excitations in the two regions meet at the transition point.
\end{document}